\documentstyle[titlepage,twoside,fleqn,12pt,epsf]{article}
\newcommand{\postscript}[2]
 {\setlength{\epsfxsize}{#2\hsize}
  \centerline{\epsfbox{#1}}}
\def\ref#1{\par\noindent \hangindent=0.4in \hangafter=1 #1 \par}
\def\eqalign#1{\null\,\vcenter{\openup\jot \m@th
  \ialign{\strut\hfill$\displaystyle{##}$&$
     \displaystyle{{}##}$\hfill \crcr#1\crcr}}\,}
\def\tempest%
{\begin{array}{ccc}
1 & 1 & 1 \\
1 & 1 & 1 \\
4 & 3 & 8
\end{array}}

\begin{document}
\centerline{}
\centerline{}
\centerline{} 
\centerline{}
\centerline{\Large Statistical Determination of}
\centerline{\Large the MACHO Mass Spectrum}
\bigskip
\centerline{\bf Cheongho Han}
\centerline{\bf Andrew Gould \footnote{Alfred P.\ Sloan Foundation Fellow}}
\bigskip
\centerline{Dept.\ of Astronomy,
The Ohio State University, Columbus, OH 43210}
\smallskip
\centerline{ cheongho@payne.mps.ohio-state.edu}
\centerline{ gould@payne.mps.ohio-state.edu}
\bigskip
\bigskip
\bigskip
\bigskip
 
\centerline{\bf Abstract}

The mass function of 51 Massive Compact Objects (MACHOs) detected 
toward the Galactic bulge is statistically estimated from 
Einstein ring crossing times $t_{\rm e}$.
For a Gaussian mass function, the best fitting parameters are 
$\langle \log (m/M_{\odot}) \rangle = -1.12$ and 
$\sigma_{\log (m/M_{\odot})} = 0.57$.
If the mass function follows a power-law distribution, the best fitting
mass cut-off and power are $m_{\rm cut} = 0.04\ M_{\odot}$ and 
$p = -2.1$.
Both mass spectra indicate that a significant fraction of events are 
caused by MACHOs in substellar mass range.
Both best determined mass functions are compared with that 
obtained from Hubble Space Telescope (HST) observations.
The power law is marginally favored ($2.3\sigma$) over the Gaussian 
mass function, and significantly over HST mass function ($5.5\sigma$),
indicating that MACHOs have a different mass function from stars 
in the solar neighborhood.
In addition, the fact that all the models have very poor 
fits to the longest four events with $t_{e} \geq 70\ {\rm days}$ 
remains a puzzle.

\bigskip
\noindent
{\it Subject headings}: gravitational lensing - dark matter - 
stars: masses

\bigskip
\bigskip
\bigskip
\bigskip
\smallskip
\centerline{submitted to {\it The Astrophysical Journal}: Jan 13, 1996}
\centerline{Preprint: OSU-TA-2/96}
 
\newpage

\section{Introduction}

Major efforts to detect MACHOs by observing microlensing 
events of source stars located in the Galactic bulge have been carried 
out by the OGLE (Udalski et al.\ 1994) and MACHO (Alcock et al.\ 1995a) 
groups.
Current and prospective observations can constrain the disk/bulge 
normalization (Stanek 1995) and the mass distribution of the bulge 
(Kiraga \& Paczy\'nski 1994; Evans 1994; Han \& Gould 1995b).

     However, it is very difficult to obtain information about 
the physical parameters of the individual lenses.
This is because the only measurable quantity from current observations, 
the Einstein ring crossing time $t_{\rm e}$, depends on a combination 
of physical parameters of the individual lenses.
The Einstein ring crossing time is related to the physical parameters 
of the lenses by
$$
t_{\rm e} = {r_{\rm e} \over v };\  r_{\rm e} = 
{(4GmD_{\rm ol}D_{\rm ls}/D_{\rm os})^{1/2} \over c},
\eqno(1.1)
$$
where $m$ and $v$ are the mass and the transverse speed of the lens, 
$r_{\rm e} $ is the Einstein ring radius, and 
$D_{\rm ol}$, $D_{\rm os}$, and $D_{\rm ls}$ 
are the distances between the observer, lens, and source.

Several ideas have been proposed to break the degeneracy of $t_{\rm e}$. 
The distance to a lens can be measured when the lensing star is bright enough 
to be detected (Kamionkowski 1994; Buchalter, Kamionkowski, \& Rich 1995). 
The ambitious and promising idea of measuring MACHO parallaxes from a 
satellite would clarify the dynamical motions of MACHOs 
(Gould 1994b, 1995; Han \& Gould 1994a).
The MACHO proper motion, $\mu = v / D_{\rm ol}$, can be measured 
photometrically (Gould 1994a; Nemiroff \& Wickramasinghe 1994;
Witt \& Mao 1994; Witt 1995; Loeb \& Sasselov 1995; Gould \& Welch 1996) 
and spectroscopically (Maoz \& Gould 1994). 
Han, Narayanan, \& Gould (1996) also proposed a method to measure $\mu$ 
by using lunar occultation.
When the information from proper motion and parallax measurements 
are combined,  the degeneracy of $t_{\rm e}$ can be completely broken, 
yielding $m$, $v$, and $D_{\rm ol}$. 
However, measuring the distance or proper motion of the lens star 
is possible only under very restricted conditions. 
Most lenses are likely to be too faint to be detected. 
A proper motion is measurable only when a lens crosses over or very close to
the face of the source star.
The Lunar occultation method is applicable only to a small number of events 
because the lunar path is restricted.
Although satellite-based parallaxes are promising, actual measurements 
are many years off.

     However, one can still obtain much information about the mass 
spectrum of MACHOs from the data on time scales that 
are provided by current observations: MACHO (Alcock, et al.\ 1995b) 
and OGLE (Udalski et al.\ 1994) have detected $> 50$ events during 1993 
bulge season.
De R\'ujula, Jetzer, \& Mass\'o (1991) have developed a method of 
``mass moments'' 
to analyze MACHO masses and Jetzer (1994) has applied this to measure 
the mean mass $(0.28\ M_{\odot})$ of bulge MACHOs using 4 OGLE events.

In this paper, we use maximum likelihood to estimate the mass function of 
microlenses which are detected toward the Galactic bulge by combining 
the latest $t_{\rm e}$ data from MACHO and OGLE with plausible models 
for the velocity and spatial distributions of the lenses and sources. 
We test both Gaussian and power-law mass functions,
and find $\langle \log (m/M_{\odot}) \rangle = -1.12$ and 
$\sigma_{\log (m/M_{\odot})} = 0.57$ for a Gaussian mass function, 
and $m_{\rm cut} = 0.04\ M_{\odot}$ and $p = -2.1$ for a power law. 
This implies that a significant fraction of events are caused by 
MACHOs in the substellar mass range.
The determined best fitting mass functions are compared with the 
observed stellar mass function as determined by
Gould, Bahcall, \& Flynn (1996) using the Hubble Space Telescope (HST).
Both best fitting mass spectra differ by more than $3\sigma$ levels 
from that of stars in the solar neighborhood.

\section{Models}

We adopt a bar-structured Galactic bulge with a double exponential disk. 
The Galactic bulge is modeled by a ``revised COBE'' model which is based 
on the COBE model (Dwek et al.\ 1995) except for the central part of 
the bulge.
In the inner $\sim 600\ {\rm pc}$ of the bulge, we adopt the high 
central density Kent (1992) model since the the COBE model does not 
match very well with observations in this region.
The disk is assumed to have an exponential distribution with vertical 
and radial scale heights of 
$H_{z} = 325\ {\rm pc}$ and $H_{R} = 3.5\ {\rm kpc}$, 
respectively (Bahcall 1986).
The models are discussed in detail by Han \& Gould (1995a).

The transverse velocity is defined as
$$
{\bf v} = {\bf v}_{\rm l} - \left[ {\bf v}_{\rm s}
{D_{\rm ol} \over D_{\rm os}}
+{\bf v}_{\rm o}{D_{\rm ls} \over D_{\rm os}} \right],
\eqno(2.1)
$$
where ${\bf v}_{\rm l}$, ${\bf v}_{\rm s}$, and ${\bf v}_{\rm o}$ 
are the projected velocities of the lens, the source, and the observer.
In our model, disk MACHOs are assumed to have a Gaussian velocity 
distribution represented by 
$$
f(v_{i}) \propto \exp \left[ - (v_{i} - \bar{v}_{i})^{2} / 
2 \sigma_{i}^{2}  \right];\ \ i=x,y,z,
\eqno(2.2)
$$
where the coordinates $(x,y,z)$ have their center at the Galactic center 
and $x$ and $z$ axes point to the Sun and the north Galactic pole, 
respectively. 
We adopt the disk component velocity distribution as 
$\bar{v}_{z,{\rm disk}} = 0\ {\rm km\ s^{-1}}$,
$\sigma_{z,{\rm disk}} = 20\ {\rm km\ s^{-1}}$, 
$\bar{v}_{y,{\rm disk}} = 220\ {\rm km\ s^{-1}}$, and
$\sigma_{y,{\rm disk}} = 30\ {\rm km\ s^{-1}}$.
The velocity of the barred bulge model is deduced from the tensor 
virial theorem (Binney \& Tremaine 1987; Han \& Gould 1995b) with 
resulting velocity dispersion components 
$(\sigma_{x'},\sigma_{y'},\sigma_{z'}) = (115.7,90.0,78.6)\ 
{\rm km\ s^{-1}}$.
Here, the coordinates $(x',y',z')$  are aligned along the axes of the 
triaxial Galactic bulge; the longest axis to the $x'$ direction and 
shortest to the $z'$ direction.
The projected velocity dispersions are then computed by
$ \sigma_{x}^{2} = \sigma_{x'}^{2} \cos^{2} \theta + 
\sigma_{y'}^{2} \sin^{2} \theta,$ 
$ \sigma_{y}^{2} = \sigma_{x'}^{2} \sin^{2} \theta + 
\sigma_{y'}^{2} \cos^{2} \theta,$ 
and $ \sigma_{z} = \sigma_{z'}$,
where $\theta \sim 20^{\circ}$ is the angle at which one views the 
major axis of the bulge.
The resulting values are $(\sigma_{x},\sigma_{y},\sigma_{z}) = 
(113.0,93.0,78.6)\ {\rm km\ s^{-1}}$.
We assume a nonrotating bulge, but see Blum (1995).

\section{Mass Spectrum}

     We have so far described the models of  ${\bf v}$
and $D \equiv D_{\rm ol}D_{\rm ls}/D_{\rm os}$, 
which are the two parameters required for the 
complete determination of the time scale $t_{\rm e}$ for a given mass $m$.
The next step is to model the mass spectrum of lenses. 
We test three mass functions.
In the first model, we assume that $\log m$ is Gaussian distributed:
$$
f(\log m) = {1 \over \sqrt{2\pi \sigma^{2}_{\log m}} } \exp \left[ 
- {(\log m - \langle \log m \rangle )^{2} \over 2 \sigma_{\log m}^{2} } \right],
\eqno(3.1)
$$
where $m$ is expressed in units of $M_{\odot}$.
The two free parameters in this mass function are the mean and 
standard deviation of the logarithmic mass: 
$\langle \log m \rangle $ and $\sigma_{\log m}$.
Note that $\sigma_{\log m}$ is the width of the mass function not the 
error estimate in the mass range of the mean.
We also test the standard power-law mass function,
$$
f(m) = K m^{p} \Theta (m-m_{\rm cut}),
\eqno(3.2)
$$
where $\Theta$ is the Heavyside step function, and 
$K \equiv (p+1)m_{\rm cut}^{-p-1}$.
Two different expressions for the mass function $f(m)$ and 
$f(\log m)$ are related by $ f(m) dm = f(\log m) d \log m$.
In addition, the stellar mass function based on observations with the 
HST (hereafter referred to as the Hubble model) 
is tested and compared with the power-law and Gaussian mass functions.
The details of the Hubble model are described in \S 4.

First, we compute the distribution of time scales $f(t'_{\rm e})$ for 
a fixed value of mass (e.g.,\ $1\ M_{\odot}$). 
To find $f(t'_{\rm e})$, one should weight events by the transverse 
speed $v$ and the cross-section (i.e.,\  Einstein ring radius 
$r_{\rm e})$.
This is because events with faster transverse speeds and larger 
cross-sections are more likely to occur.
Then the distribution function of $t'_{\rm e}$ is computed by 
$$
f(t'_{\rm e}) = \xi \int_{0}^{d_{\rm max}} 
dD_{\rm os}n(D_{\rm os}) \int _{0}^{D_{\rm ol}} dD_{\rm ol}
\rho(D_{\rm ol}) D^{1/2}
$$
$$
\times \int dv_{y} \int dv_{z} v f(v_{y}, v_{z}) 
\delta \left( t'_{\rm e} - \sqrt{4GM_{\odot} D} /c v \right), 
\eqno(3.3)
$$
where $D \equiv D_{\rm ol}D_{\rm ls} / D_{\rm os} $, and $\xi$ is a
normalization factor to be discussed below.
We assume that the bulge is cut off at $4\ {\rm kpc}$ from the Galactic 
center (at $R_0=8\ {\rm kpc}$), and therefore $d_{\rm max} = 12\ {\rm kpc}$.
The increase of volume element, and thus increase in total number of 
stars in the volume is assumed to be compensated by the decrease of 
observable stars due to decreasing detectability with distance.
This is equivalent to the parameter $\beta = -1$ model of 
Kiraga \& Paczy\'nski (1994).
We take disk self-lensing events into consideration by setting the 
lower limit of source stars to be 0 in equation (3.3).

    At the second stage, we compute the actual time scale distribution 
$f(t_{\rm e})$ of MACHOs whose masses are distributed by a function 
$\phi(m)$. 
The distribution $f(t_{\rm e})$ is obtained by convolving two functions, 
$f(t'_{\rm e})$ and $\phi (m)$:
$$
f(t_{\rm e}) = \eta (t_{\rm e}) \int_{m_{\rm cut}}^{m_{\rm up}} 
dm m^{1/2} \phi (m) \int_{0}^{\infty} dt'_{\rm e} 
f(t'_{\rm e}) \delta(t_{\rm e} -t'_{\rm e}\sqrt{m}),
\eqno(3.4)
$$
where $\eta(t_{\rm e})$ is the detection efficiency, 
and the factor $m^{1/2}$ is included to weight events by their cross 
section which is $r_{\rm e} \propto m^{1/2}$.
In the computation of $f(t_{\rm e})$, we assume an upper mass limit 
$m_{\rm up} = 10\ M_{\odot}$. 
Using the assumed distributions of $v$, $D$, and $m$, we find the 
best fitting parameters of the mass functions; 
$\langle \log m \rangle$ and $\sigma_{\log m}$ for the Gaussian and 
$m_{\rm cut}$ and $p$ for the power-law mass spectrum.
The best fitting mass function is obtained by comparing model and observed 
time scale distributions.
For this, we use the maximum likelihood test in which the statistic 
is computed by
$$
{\rm ln\ L} = \sum_{i=1}^{N_{\rm tot}} {\rm ln}f(t_{{\rm e,obs},i}),
\eqno(3.5)
$$
where $t_{\rm e,obs}$ is the observed time scale.
Prior to the computation of ln L, the test distribution $f(t_{\rm e})$ 
constructed from equations (3.3) and (3.4) is corrected by the detection 
efficiency $\eta (t_{\rm e})$ using the functions provided by 
Alcock et al.\ (1995b) for MACHO and Zhao, Spergel, \& Rich (1994) 
for OGLE.
Since the efficiencies for each group are different, 
the likelihood has been computed separately and summed:
$$
{\rm ln\ L} = \sum_{j} \sum_{i=1}^{N_{j}} {\rm ln}f_{j}(t_{{\rm e,obs},i}),
\eqno(3.6)
$$
where the subscripts $j={1,2}$ represent values of MACHO and OGLE, 
respectively.
In this test, one determines the best fitting distribution that 
maximizes ln L.
The likelihood statistic is related to the uncertainty by 
${\mit\Delta} ({\rm ln\ L}) = \sigma^{2} / 2$. 
Since our concern is finding best fitting mass spectrum not the total 
amount of optical depth or frequency, we leave the overall normalization 
$\xi$ as a variable so that the expected number of events matches 
with that of the observed events.

The distribution of observed time scale
is obtained from data measured by MACHO (Alcock et al.\ 1995b) and
OGLE (Udalski et al.\ 1994) groups.
Out of 55 candidate events detected (45 by MACHO and 12 by OGLE,
with 2 overlapping ones),
we exclude four suspected by MACHO (Alcock et al.\ 1995b) as being
either variables or influenced by neighboring variables.
For the photometry toward a dense field such as Baade's Window, 
it is expected that blending of stars causes errors in 
determining time scales.
The blending is more serious for faint stars and resultant time scales 
determined  are shorter than the actual values.
To check how seriously the blending affects the time scale measurement, 
we plot the time scale versus $V$ mag of source stars in Figure 1.
In the figure, data from MACHO and OGLE groups are marked by 
filled and empty circles, respectively.
Except three long events detected by MACHO group, there does not 
exist any trend in the figure. 
We therefore ignore blending in our analysis.

\section{Result and Discussion}

The results of the ln L computation in parameter space are shown 
as contour maps in Figure 2 for power-law (upper panel) and Gaussian 
(lower panel) mass functions, respectively.
In the figure, the best fitting positions in parameter space are 
marked with `x' and the contours are drawn at $1\sigma$, $2\sigma$, 
and $3\sigma$ levels; the power-law fit is better than the 
Gaussian by $\sim 2.3\ \sigma$ and better 
than the Hubble by $\sim 5.3\ \sigma$.
For the Gaussian mass function, we find the best fitting parameters of 
$\langle \log m \rangle = -1.12$ and $\sigma_{\log m} = 0.57$.
The best fitting parameters for the power-law distribution are 
$m_{\rm cut} =  0.04\ M_{\odot}$ and $p = -2.1$. 
Therefore, a significant fraction of events might be caused by 
lenses with mass less than the hydrogen-burning mass limit 
($\sim 0.1\ M_{\odot}$) regardless of the assumed form 
of the mass spectrum.

The determined best fitting mass functions are compared with that  
of local disk stars to check similarities and differences between 
the two populations.
Gould et al.\ (1996) used HST to measure the luminosity 
function of local Galactic disk stars.
They then applied mass-luminosity relation from Henry \& McCarthy 
(1990) to obtain a local mass function.
The Hubble mass function is shown and compared with the best fitting  
power-law and Gaussian mass functions in Figure 3.
We have added the white dwarf (WD) population to the Hubble model,
the bumpy peak at $\log m \sim -0.3$, 
which is modeled based on the observation of McMahan (1989).
The mass function of WD is normalized using the mass density 
$\rho_{\rm WD} \sim 0.005\ M_{\odot} {\rm pc^{-3}}$ (Bahcall 1984).
Luminous massive stars would have a low scale height and thus only 
nearby objects could contribute to microlensing events seen toward 
Baade's Window, $b \sim -4^{\circ}$. 
Then it is expected one would detect these objects due to both 
their closeness and high luminosity.
However, there are no events in which the lens is bright enough to 
be easily detected.
Therefore, we cut off the Hubble mass function at the upper limit of 
$\sim 2.2\ M_{\odot}$.
The best fitting time scale distributions $f(t_{\rm e})$ for the 
three mass functions are shown in Figure 4 and are compared with 
the observed distribution represented by histograms in each figure.
Both power-law and Gaussian mass functions fit significantly 
better than the HST model does, implying that the lenses have very 
different mass distribution from stars in the Galactic disk.

It is puzzling that any type of assumed mass function cannot 
match very well for the long time-scale events 
($t_{\rm e} \geq 70\ {\rm days}$). 
Both observed and best fitting $f(t_{\rm e})$ are magnified 
and shown in the smaller boxes inside each panel so that the 
differences between observation and the best fits of each mass 
functions can be compared more easily in the long event region.
The expected number of events with $t_{\rm e} \geq 70$ days are 
$0.68$ and $1.00$, for the power-law and Hubble mass functions, 
respectively.
Thus, the Poisson probabilities of observing 4 or more long events are 
$\sim 1\%$ for the power-law and $\sim 2\%$ for the Hubble mass function.
The probability is even lower (0.55 events) for the Gaussian.
Even though only 4 events have time scale longer than 70 days, 
these long events are important because they are responsible for the 
significant fraction of optical depth measured toward the bulge.

Long time-scale events can be produced under various conditions.
First, the long events might be caused by a mass function in which a 
larger fraction of lenses are in the higher mass range.
If the mass function is a power-law, this can be achieved by a higher 
$m_{\rm cut}$ and lower $p$.
In Figure 5, the expected time-scale distributions for various lower 
values of power (upper panel) and higher mass cut-off (lower panel) 
are shown and compared to that of best fitting distribution.
However, as one or both of the conditions are satisfied to explain 
the long events, the fit deviates seriously in the low mass region 
where the majority of events are located.
Therefore, the observed bimodal pattern of $f_{\rm obs}(t_{\rm e})$ 
is not likely to be due to a lower power or higher mass cut-off 
than our determination.
Another explanation would be a MACHO population with low transverse speed. 
For example, if the velocity dispersion of bulge population is very 
low, a MACHO will cross slowly over the Einstein ring, thus producing 
very long events.
However, this explanation does not seem to be plausible because
the velocity dispersion of the bulge is observationally 
reasonably constrained, and thus it cannot be arbitrarily low.
Another possibility is that longer events might be caused if a lens 
is composed of a binary system which is so closely spaced that one 
would detect it as a single lens instead of a binary lens.
If the fraction of close binary systems is significant, the observed 
time scale distribution would be significantly different from that 
expected with an assumed single star mass distribution.
However, the fraction of binary stars whose separations are less 
than 2 AU, which is a typical Einstein ring radius, is $< 25 \%$ for 
G type stars in the solar neighborhood (Duquennoy \& Mayor 1991).
Therefore, in the computation we do not take into consideration the 
modification of the time scale distribution due to binary systems. 
Finally, another explanation would be a bimodal mass distribution with 
the second population composed of heavy dark MACHOs 
(e.g., black holes, neutron stars, or white dwarfs) 
with low transverse velocity.
If, for example, there was a kinematically cold population of these 
objects, they would have a low scale height and hence would be 
observed mostly near the sun for sources near Baade's Window, i.e., 
a few degrees from the Galactic plane.
Such a population would then have a very low transverse speed.

It is curious that the best fitting power-law does not deviate much from
 that of the Salpeter mass function, which is valid in the mass 
range $M > 1\ M_{\odot}$.
Determination of the mass function in the solar neighborhood by 
Miller \& Scalo (1979) indicated that the increase in the number of 
stars becomes shallower (decreasing $p$) at lower mass than the classical 
Salpeter estimate of $p \sim -2.35$ for stars with  $m > 1\  M_{\odot}$.
Their estimate is $p \sim -1.4$ in the mass range 
$0.1\ M_{\odot} \leq m \leq 1\ M_{\odot}$.
By contrast, our determined power $p=-2.1$ happen to be close to the 
classical value, and would seem to imply that the classical 
Salpeter mass function extends to lower mass objects.
Note, however, that the Hubble function, in which
the number of stars decreases as mass decreases after passing the 
maximum at $m \sim 0.5\ M_{\odot}$, is inconsistent with such a power law.

\section{Conclusion}

We determine the mass spectrum of MACHO detected toward the Galactic bulge
by modeling velocity and matter distributions.
For the best fitting power-law and Gaussian mass functions, 
a significant fraction of events is caused by MACHOs in the 
substellar mass range. 
In addition, the mass spectrum seems to differ significantly 
from that of local Galactic disk stars.
However, caution is required in interpreting these results since 
the mass spectrum of MACHOs determined here is based on models
of density and velocity distributions.
The parameters of these models, e.g., radial and vertical 
scale height of the disk, total mass of the 
Galactic bulge, and velocity field in the central Galactic bulge,
are not unambiguously determined.
Different models can lead to different conclusion.
For illustration, we arbitrarily set the bulge velocity 
dispersions at high levels of 
$\sigma_{x'}=\sigma_{y'}=\sigma_{z'} = 100\ {\rm km}\ {\rm s}^{-1}$.
The best fitting Gaussian mass spectrum is obtained the same way
as described in \S 3 and the best fitting parameters are marked
by $\odot$ in Figure 2.
However, the differences are small: 
${\mit\Delta}\langle \log m \rangle = -0.138$ and
${\mit\Delta} \sigma_{\log m} =0.056$.
An additional uncertainty comes from the low detection efficiency 
for very short events.
It is possible that a class of very low mass MACHOs is generating 
very short ($t_{\rm e} \leq 3\ {\rm days}$) events that go 
completely undetected.
We have issued this possibility in this paper.
The MACHO group is presently conducting a ``spike analysis'' to 
determine whether such ultra-short events in fact occur.

{\bf Acknowledgement}: We would like to thank M. Everett for 
making very helpful comments and suggestions. 
This work is supported by a grant AST 9420746 from the NSF.

\newpage
\vskip50mm
\centerline{\bf REFERENCE}
\bigskip

%\ref{Alcock, C., Allsman, R.\ A., Alves, D., Axelrod, T.\ S.,
%Bennett, D.\ P., Cook, K.\ H., Freeman, K.\ C., Griest, K.,
%Guern, J., Lehner, M.\ J., Lehner, M.\ J., Marshall, S.\ L.,
%Peterson, B.\ A., Pratt, M.\ R., Quinn, P.\ J., Rogers, A.\ W.,
%Stubbs, C.\ W., \& Sutherland, W.\  1995a, ApJL, submitted}
%\ref{Alcock, C., Allsman, R.\ A., Axelrod, T.\ S., Bennett, D.\ P., 
%Cook, K.\ H., Freeman, K.\ C., Griest, K., Guern, J., 
%Lehner, M.\ J., Marshall, S.\ L., Park, H.\ S., Perlmutter, S., 
%Peterson, B.\ A., Pratt, M.\ R., Quinn, P.\ J., Rodgers, A.\ W., 
%Stubbs, C.\ W., Sutherland, W.\ 1995b, preprint, astro-ph/9512146}
\ref{Alcock, C.\ et al.\ 1995a, ApJL, submitted}
\ref{Alcock, C.\ et al.\ 1995b, ApJ, submitted}
\ref{Bahcall, J. N.\ 1984, ApJ, 276, 169}
\ref{Bahcall, J. N.\ 1986, ARA\&A, 24, 577}
\ref{Binney, J., \& Tremaine, S.\ 1987, Galactic Dynamics (Princeton 
University Press, Princeton), 67}
\ref{Blum, R.\ D.\  1995, ApJ, 444L, 89}
\ref{Buchalter, A., Kamionkowski, M., \& Rich, R.\ M.\ 1995, preprint, 
CU-TP-715}
\ref{De R\'ujula, A., Jetzer, P., \& Mass\'o, E.\ 1991, MNRAS, 250, 348}
\ref{Duquennoy, A., \& Mayor, M.\ 1991, A\&A, 248, 485}
%\ref{Dwek, E., Arendt, R. G., Hauser, M. G., Kelsall, T., Lisse, C. M.,
%Moseley, S. H., Silverberg, R. F., Sodroski, T. J., \& Weiland, J.\ 1995, ApJ, 
%445, 716}
\ref{Dwek, E.\ et al.\ 1995, ApJ, 445, 716}
\ref{Evans, N. W.\ 1995, ApJ, 437, L31}
\ref{Gould, A.\ 1994a, ApJ, 421, L75}
\ref{Gould, A.\ 1994b, ApJ, 421, L71}
\ref{Gould, A.\ 1995, ApJ, 441, L21}
\ref{Gould, A., Miralda-Escud\'e, J., \& Bahcall, J. N.\ 1994, ApJ, 423, L105  }
\ref{Gould, A., Bahcall, J. N., Flynn, C.\ 1996, ApJ, submitted}
\ref{Gould, A., \& Welch, D.\ 1996,ApJ, 464, 000}
\ref{Han, C., \& Gould, A.\ 1995a, ApJ, 447, 53}
\ref{Han, C., \& Gould, A.\ 1995b, ApJ, 449, 521}
\ref{Han, C., Narayanan, V., \& Gould, A.\ 1996, ApJ, 461, 000}
\ref{Henry, T. J., \& McCarthy, D. W. 1990, ApJ, 350, 334}
\ref{Jetzer, P.\ 1994, ApJ, 432, L43}
\ref{Kamionkowski, M.\ 1995, ApJ, 442, L9} 
\ref{Kent, S. M.\ 1992, ApJ, 387, 181}
\ref{Kiraga, M., \& Paczy\'nski, B.\ 1994, ApJ, 430, L101}
%\ref{Liebert, J., \& Probst, R. G.\ 1987, ARAA, 25, 473}
\ref{Loeb, A., \& Sasselov, D.\ 1995, ApJ, 449, L33}
\ref{Maoz, D. \& Gould, A.\ 1994, ApJ, 425, L67}
\ref{McMahan, R. K.\ 1989, ApJ, 336, 409}
\ref{Miller, G. E. \& Scalo, J. M.\ 1979, ApJS, 41, 513}
\ref{Nemiroff, R. J. \& Wickramasinghe, W. A. D. T.\  1994, ApJ, 424, L21}
\ref{Stanek, K. Z.\ 1995, ApJ, 441, L29}
%\ref{Udalski, A., Szyma\'nski, M., Stanek, K. Z., Kalu\.zny, J., Kubiak, M.,
%Mateo, M., Krzemi\'nski, B., \& Venkat, R.\ 1994, Acta Astron., 44, 165}
\ref{Udalski, A.\ et al.\ 1994, Acta Aston., 44, 165}
\ref{Witt, H.\ J.\ 1995, ApJ, 449, 42}
\ref{Witt, H.\ J., \& Mao, S.\ 1994, ApJ, 429, 66}
\ref{Zhao, H., Spergel, D. N., \& Rich, R. M.\ 1995, ApJ, 440, L13}

\begin{figure}
\postscript{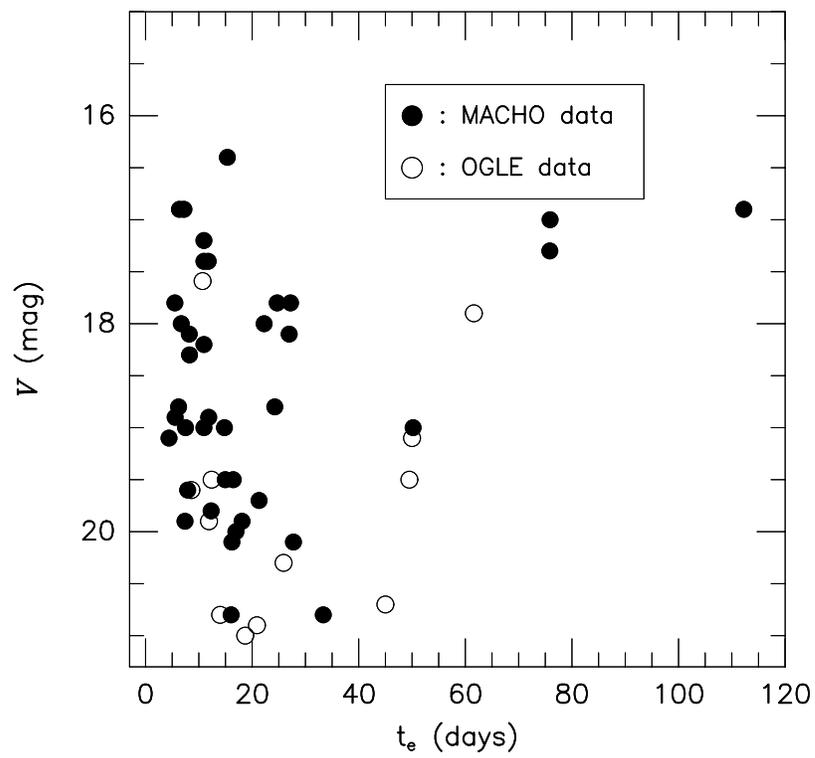}{1.0}
\caption{
Einstein crossing times of lensing events obtained 
by MACHO and OGLE versus $V$ mag of source stars.
}
\label{fig1}
\end{figure}

\begin{figure}
\postscript{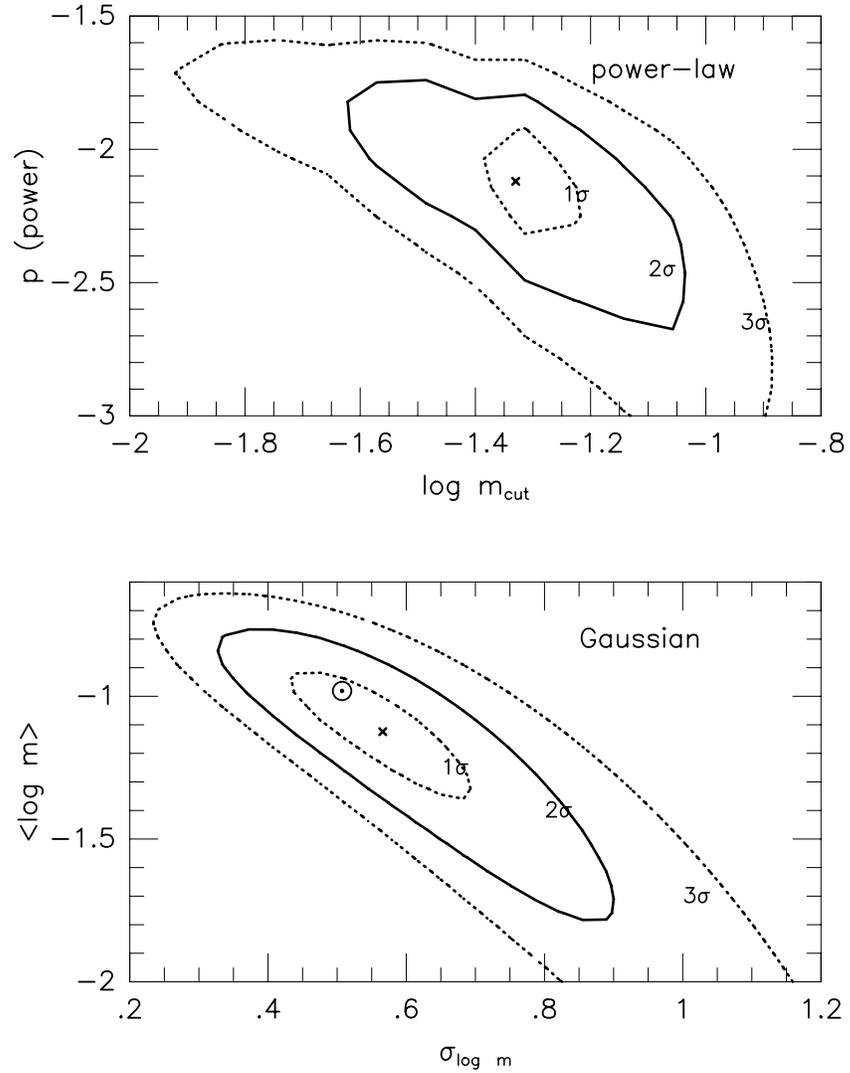}{1.0}
\caption{
Likelihood contour for the power-law and Gaussian mass functions.
The contours are drawn at $1\sigma$, $2\sigma$, and $3\sigma$ 
levels from the best fitting position marked by `x'.
Th best fitting point with higher bulge velocity dispersion,
$\sigma=100\ {\rm km}\ {\rm s}^{-1}$, is marked by $\odot$ 
(see \S\ 5).
}
\label{fig2}
\end{figure}

\begin{figure}
\postscript{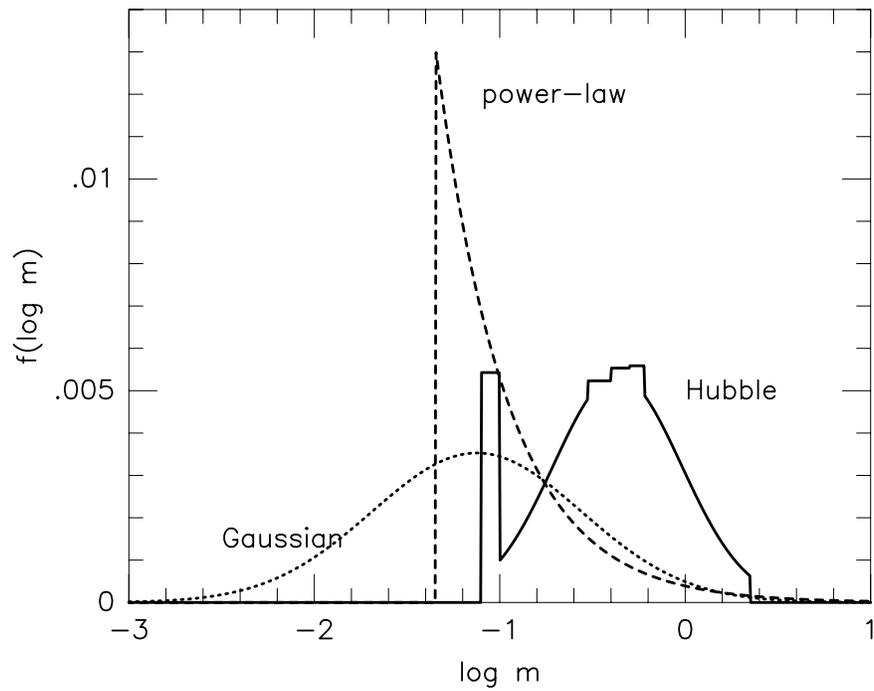}{1.0}
\caption{
The best fitting mass spectrum for the power-law and Gaussian 
mass functions are compared with the Hubble mass function.
}
\label{fig3}
\end{figure}

\begin{figure}
\postscript{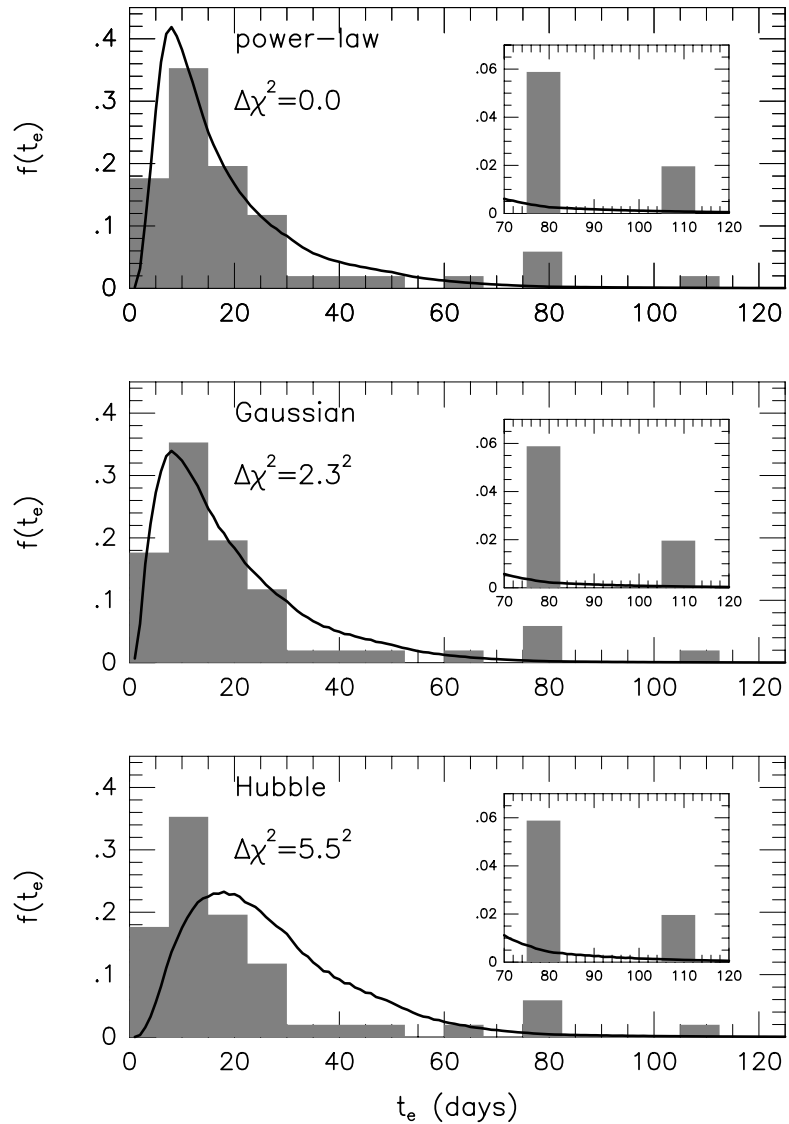}{1.0}
\caption{
The respective best fitting time scale distribution functions 
$f(t_{\rm e})$ for power-law, Gaussian, and Hubble mass functions.
They are compared with the observed distribution, $f_{\rm obs}(t_{\rm e})$, 
represented by a histogram.
The distribution in the long time scale region $(t_{\rm e} \geq 70\ {\rm days})$
are rescaled and drawn in the smaller boxes in each panel.
}
\label{fig4}
\end{figure}

\begin{figure}
\postscript{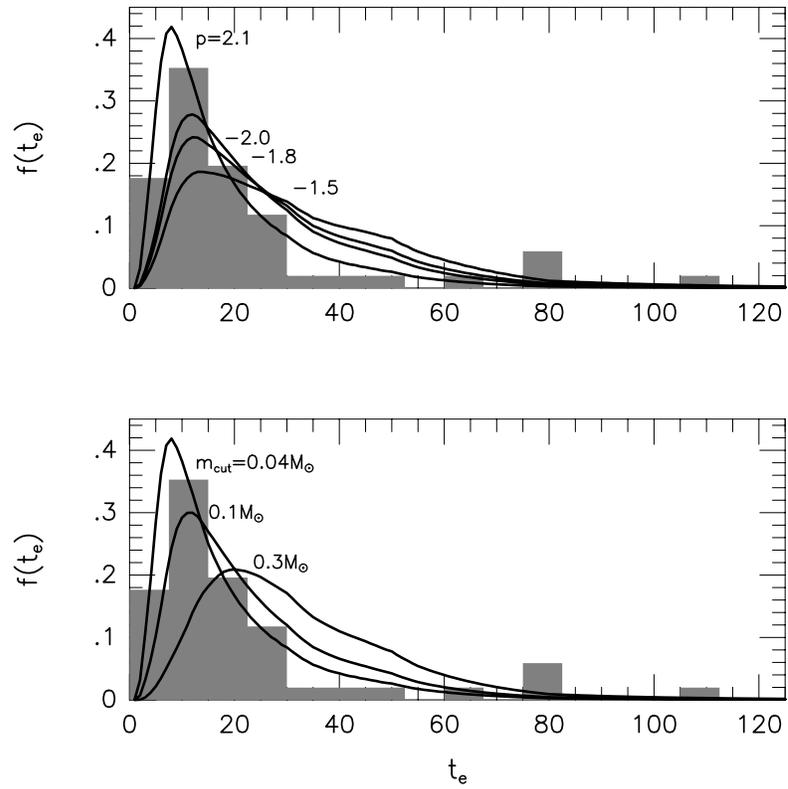}{1.0}
\caption{
The time scale distribution functions for lower value of power $p$ and the 
higher mass cutoff $m_{\rm cut}$ are compared with the best fitting
distribution with $p=-2.1$ and $m_{\rm cut}=0.04\ M_{\odot}$.
}
\label{fig5}
\end{figure}

\end{document}